\documentclass[12pt,preprint]{aastex}
\usepackage[left=1in,top=1in,bottom=1in,right=1in,nohead,nofoot]{geometry}
\def\e{{\rm E}}
\def\hbn{{\hfil\break\noindent}}
\begin{document}

\begin{center}

{\bf \Large The Demographics of Extrasolar Planets Beyond the Snow Line with Ground-based Microlensing Surveys}

\bigskip
White Paper for the Astro2010 PSF Science Frontier Panel\\
\bigskip
B. Scott Gaudi\\
The Ohio State University\\
gaudi@astronomy.ohio-state.edu\\
614-292-1914\\
\bigskip
J.P.\ Beaulieu$^1$, David P.\ Bennett$^2$, Ian A.\ Bond$^3$, Subo Dong$^4$, Andrew Gould$^4$, Cheongho Han$^5$,
Byeong-Gon Park$^6$, 
Takahiro Sumi$^8$\\
\bigskip
$^1$Institute d'Astrophysique de Paris\\
$^2$University of Notre Dame\\
$^3$Massey University\\
$^4$The Ohio State University\\
$^5$Chungbuk National University\\
$^6$Korea Astronomy and Space Science Institute\\
$^7$Nagoya University
\end{center}

\newpage

{\bf 1. Overview}

In the currently-favored paradigm of planet formation, the location of
the snow line in the protoplanetary disk plays a crucial role (e.g.\
Ida \& Lin 2004).  Determining the demographics of planets with masses
of the Earth or larger beyond the snow line of stars of various masses
is thus essential for testing this model.  In the inner parts of this
region, RV probes the gas giants but not the ice giants nor, of
course, terrestrial planets.  RV cannot make reliable measurements in
the outer part of this region at all because the periods are too
long. Future astrometry missions (such as {\it SIM Lite}) could probe the
inner regions down to terrestrial masses, but are also limited by
their finite lifetime in the outer regions.  Microlensing is sensitive
to planets that are generally inaccessible to these other methods, and
in particular is most sensitive to cool planets at or beyond the snow
line, including very low-mass (i.e.\ terrestrial) planets.  Hence,
microlensing is uniquely suited and so essential for a comprehensive
study of this region.  Microlensing is also sensitive
to planets orbiting low-mass stars, free-floating planets, planets in
the Galactic bulge and disk, and even planets in external galaxies.
These planets can also provide critical constraints on models of
planet formation.

In its final report, the ExoPlanet Task Force (Lunine et al.\ 2008)
recognized the major role microlensing has to play in determining the
demographics of planets throughout the Galaxy, writing that ``[t]he
statistics of planetary masses and separations available from a
microlensing survey are vital for constraining the theory of planet
formation."  They recommended that both ground and space-based
microlensing be supported in the next 5-10 years.  Here we focus on
ground-based microlensing, and leave the discussion of space-based
surveys for a separate paper (Bennett et al.\ 2009b).

We briefly review the properties of and current results from the
microlensing method (see Bennett 2009a for a more thorough discussion),
and then outline the potential of, and progress toward, next
generation ground-based microlensing surveys.  Detailed models of such
surveys have already been carried out, and the required network of
1-2m class telescopes with wide FOV instruments is beginning to be
constructed by several countries, including Japan, New Zealand, Poland, 
and South Korea. The US still has a substantial role to play in
supporting these initiatives, through sharing of technical expertise
in the design and construction of the telescopes, analysis of
microlensing events from these surveys, contribution to follow-up
observations, and potentially monetary and/or in kind contributions to
supplement or augment the already-planned hardware investments.

{\bf 2. The Properties of Microlensing Planet Searches}

If a foreground star (``lens'') becomes closely aligned with 
a more distant star (``source''), it bends the source light 
into two images.  The resulting magnification is a monotonic
function of the projected separation. For Galactic stars, the image 
sizes and separations are of order $\mu$as and mas respectively, 
so they are generally
not resolved. Rather ``microlensing events'' are recognized from
their time-variable magnification (Paczynski 1986), which typically occurs on
timescales $t_\e$ of months, although it ranges from days to years in
extreme cases.  Presently about 900 microlensing events are
discovered each year, almost all toward the Galactic bulge.

If one of these images passes close to a planetary companion of the lens star,
it further perturbs the image and so changes the magnification.
Because the range of gravitational action scales $\propto \sqrt{M}$, where
$M$ is the mass of the lens, the planetary perturbation typically
lasts $t_p\sim t_\e\sqrt{m_p/M}$, where $m_p$ is the planet mass.
That is, $t_p\sim 1\,$day for Jupiters and $t_p\sim 1.5\,$hours for
Earths.  Hence, planets are discovered by intensive, round-the-clock
photometric monitoring of ongoing microlensing events (Mao \& Paczynski 1991, Gould \& Loeb 1992).

{\bf 2.1 Sensitivity of Microlensing}

While, in principle, microlensing can detect planets of any mass and
separation, orbiting stars of any mass and distance from Earth, 
the characteristics of microlensing favor some
regimes of parameter space.
\hbn $\bullet$ {\bf Sensitivity to Low-mass Planets}:  Compared to other techniques, microlensing
is more sensitive to low-mass planets. This is because the {\it amplitude
of the perturbation does not decline as the planet mass declines},
at least until
mass goes below that of Mars (Bennett \& Rhie 1996).  The {\it duration} does decline as 
$\sqrt{m_p}$
(so higher cadence is required for small planets) and the probability
of a perturbation also declines as $\sqrt{m_p}$ (so more stars must
be monitored), but if a signal is detected, its magnitude is typically
large ($\ga 10\%$), and so easily characterized and unambiguous.
\hbn $\bullet$ {\bf Sensitivity to Planets Beyond the Snow Line}:  Because microlensing works by
perturbing images, it is most sensitive to planets that lie at projected
distances where the images are the largest.  This so-called
``lensing zone'' lies within a factor of 1.6 of the
Einstein ring radius, 
$r_\e = \sqrt{(4 G M/c^2)D_s x(1-x)}$, 
where $x=D_l/D_s$ and  $D_l$ and $D_s$ are the distances to the lens and source. 
At the Einstein ring, the equilibrium temperature is
\begin{equation}
T_\e = T_\oplus\biggl({L\over L_\odot}\biggr)^{1/4}
\biggl({r_\e\over \rm AU}\biggr)^{-1/2}=
70\,{\rm K}\,{M\over 0.5\,M_\odot}[4x(1-x)]^{1/4},
\end{equation}
where we have adopted a simple model for lens luminosity $L\propto M^5$,
and assumed $D_s=8\,$kpc.  Hence, microlensing is
primarily sensitive to planets 
beyond the snow line.  
\hbn $\bullet$ {\bf Sensitivity to Free Floating Planets:}
Because the microlensing effect arises directly from the planet mass, 
the existence of a host star is not required for detection.  
Thus, microlensing maintains significant
sensitivity at arbitrarily large separations, and in particular
is the only method that is sensitive to old, free-floating planets.
\hbn $\bullet$ {\bf Sensitivity to Planets in the Disk and Bulge}:  Microlensing searches require dense star fields
and so are best carried out against the Galactic bulge, which is 8 kpc away.
Given that the Einstein radius peaks at $x=1/2$, it is most sensitive
to planets that are 4 kpc away, but maintains considerable sensitivity
provided the lens is at least 1 kpc from both the observer and the source.
Hence, microlensing is about equally sensitive to planets in the bulge
and disk of the Milky Way.
However, future specialized searches may also be sensitive to closer planets
and to planets in other galaxies, particularly the LMC and SMC, and M31. 
\hbn $\bullet$ {\bf Sensitivity to Planets Orbiting a Wide Range of Host Stars}:  Microlensing is about equally sensitive to
planets independent of host luminosity, i.e., planets of stars all along 
the main sequence, from G to M, as well as white dwarfs and brown dwarfs.  
By contrast, other techniques are generally challenged to detect planets
around low-luminosity hosts.
\hbn $\bullet$ {\bf Sensitivity to Multiple Planet Systems:} 
In general, the probability
of detecting two planets (even if they are present) is the square
of the probability of finding one, which means it is usually very small.
However, for high-magnification events, the planet-detection probability
is close to unity (Griest \& Safizadeh 1998), and so its square is also near unity 
(Gaudi et~al.\ 1998).  
In rare cases, microlensing can also detect exomoons (Bennett \& Rhie 2002).

{\bf 2.2 Planet and Host Star Characterization}

It has often been stated that the ability of microlensing to provide detailed information
about individual systems is very limited. This perception comes from the fact that (1) the
host stars are typically 
distant and faint, making follow-up work is difficult, (2) in the
overwhelming majority of microlensing events, the only parameter that
can be constrained that contains any information about the primary lens is
the event timescale $t_{\rm E}$, which is a degenerate combination of mass,
distance, and transverse velocity of the lens, (3) microlensing detections
routinely provide only the mass ratio of the planet and host star (Gaudi \& Gould 1997), and so the mass
of the planet is typically not known without a constraint on the primary mass, and 
furthermore (4) the only constraint on the planet orbit is $b_\perp$, 
the instantaneous angular separation between
the planet and host star at the time of the event in
units of the angular Einstein ring radius $\theta_{\rm E}=r_{\rm E}/D_l$.   Since $\theta_{\rm E}$, the inclination, phase, and eccentricity
of the orbit are all unknown, $b_\perp$ alone provides very
little information about the orbit. 

Experience has shown that, in reality, much more information
can typically be gleaned from a combination of a
detailed analysis of the light curve and high-resolution follow-up 
imaging.   First, for the majority of
planets detected via microlensing, the `smoothing' effects of the
finite source size are detectable during sharp features in the
light curve caused by the planet.  The magnitude of this effect
yields $\theta_{\rm E}$.  Second,
for many long-duration events, it is
also possible to measure
the deviations in the microlensing light curve caused by the fact that
the Earth is accelerating.  This `microlens parallax' allows one to constrain $\tilde
r_{\rm E}$, the Einstein ring radius
projected onto the observer plane.  Third, for a substantial
fraction of events, the lens light can be detected during and after
the event (often in several different filters),
allowing for a photometric estimates of the lens mass and distance,
and so estimates of planet mass $m_p$ and projected separation (Bennett et al.\ 2007).
Finally, in some cases,
the orbital motion of the planet during the microlensing event
can be detected.  Generally, if the lens mass
is known, and under the assumption of a circular
orbit, a measurement of the effects of orbital motion 
specify the full orbit of the planet (including inclination), up to a
two-fold degeneracy (Dong et al.\ 2008a).  
In some cases, additional information can be used to break this
degeneracy and strongly constrain the eccentricity of the orbit.

In many instances, several of these pieces of information can be measured
in the same event, often providing complete or even redundant measurements
of the mass, distance, and transverse velocity of the event.  For example,
a measurement of $\theta_{\rm E}$ from finite source effects, when combined
with a measurement of $\tilde r_{\rm E}$ from microlens parallax, yields
the lens mass $M=(c^2/4G)\tilde r_{\rm E} \theta_{\rm E}$, distance
$d_{l}^{-1} = \theta_{\rm E}/\tilde r_{\rm E} + d_s^{-1}$,
and transverse velocity.  

For all eight published microlensing planet detections,
it has been possible to measure $\theta_{\rm E}$ using finite source effects
and so partially break the $t_\e$ degeneracy. 
For four systems (with five planets) it has been possible to uniquely
measure the mass and distance to the planet and primary.  Follow-up
imaging with {\it HST} and/or ground-based adaptive optics 
would allow one to completely solve the remaining three events.
In one exceptional case (Gaudi et al.\ 2008) redundant information
on the primary mass and distance was obtained, and furthermore 
it was possible to constrain the eccentricity and 
inclination of the planetary orbit. 

\begin{figure}[!ht]
\plottwo{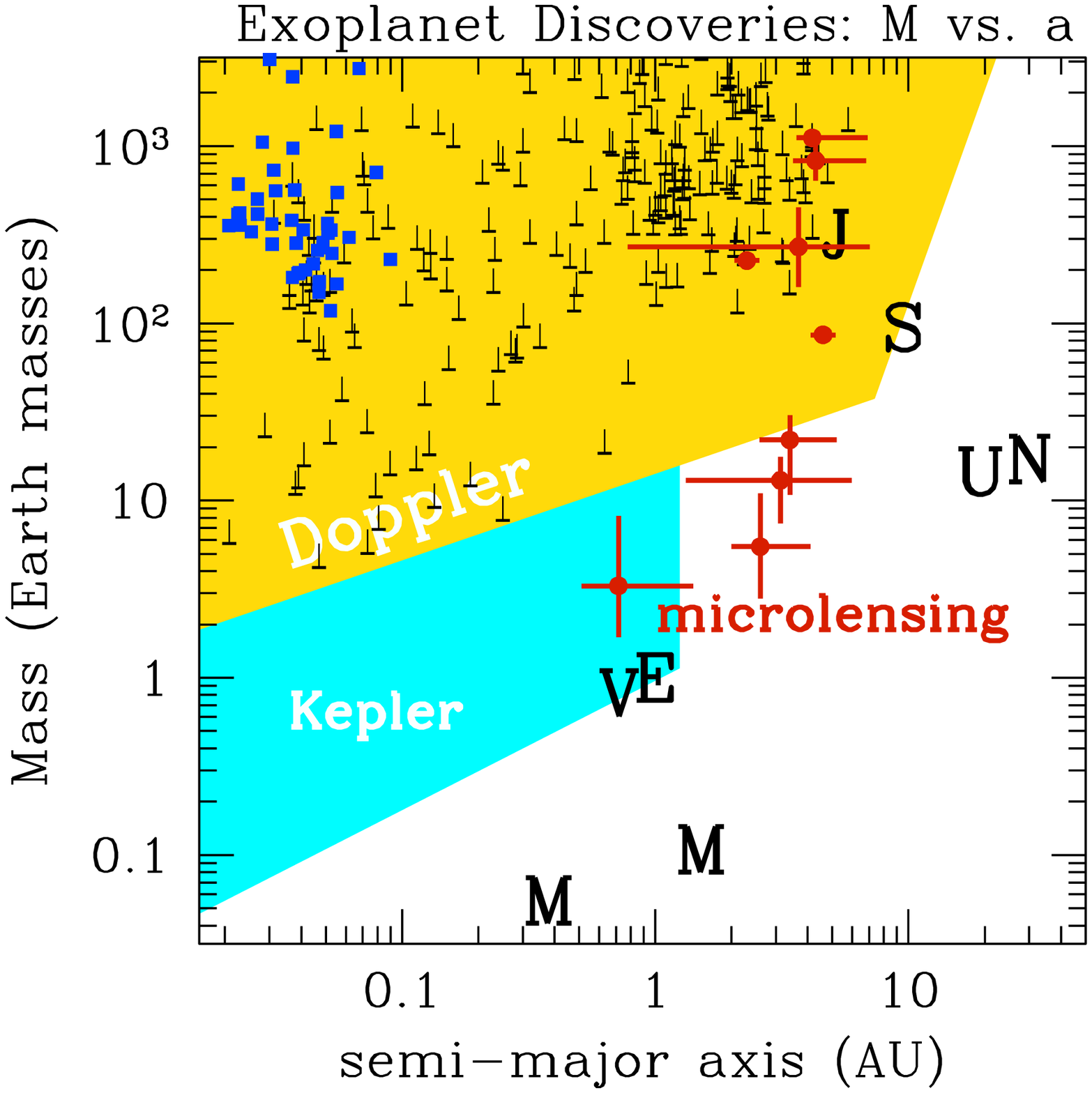}{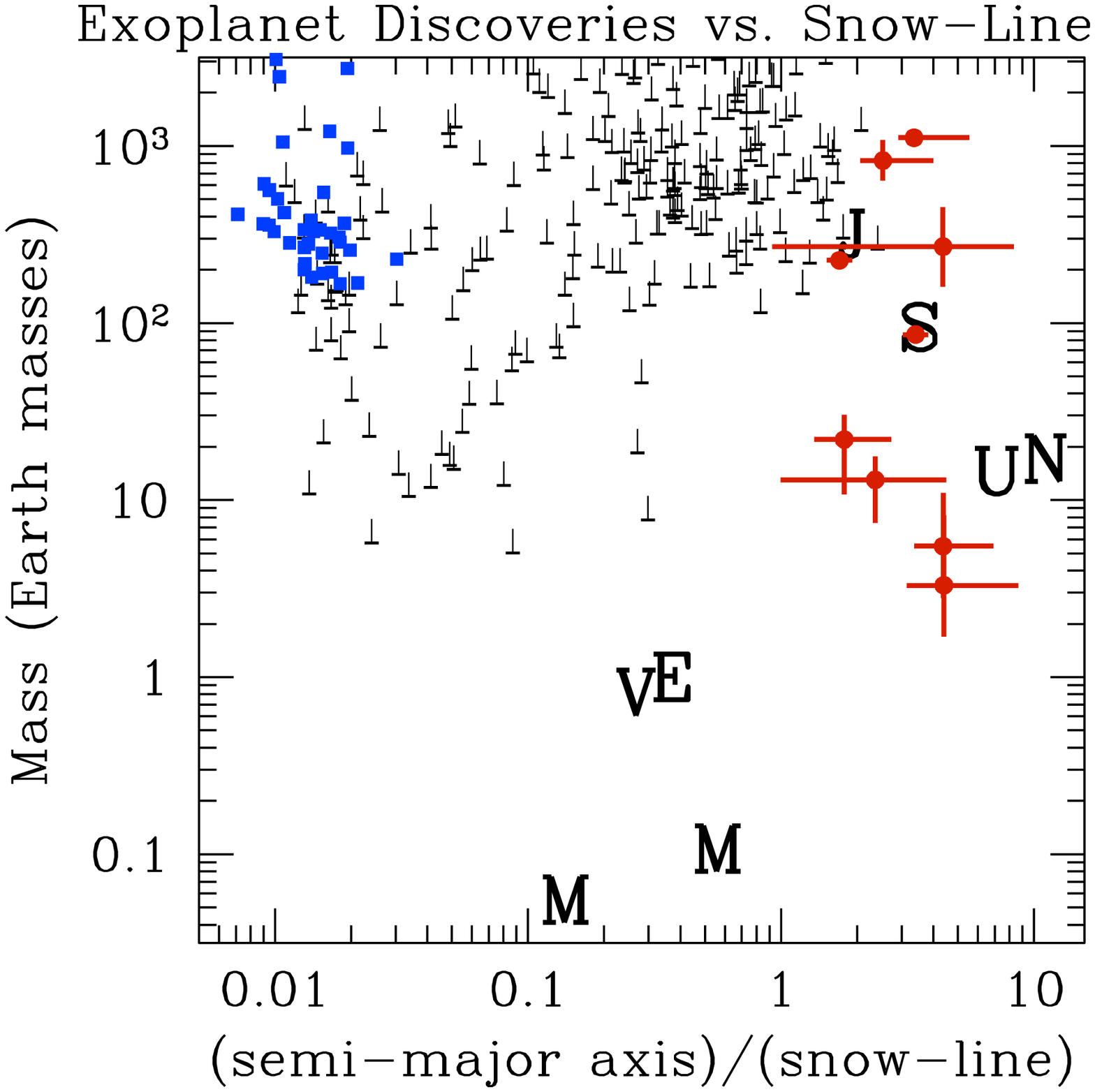}
\vskip-0.2in
\caption{ 
Distribution of known planets is plotted as functions
of mass vs.~semi-major axis (left) and
semi-major axis normalized to the location of the
snow-line (right). Planets found by the Doppler/RV
method (black lines with error bars extending
upward to indicate the $\sin i$ uncertainty), transits
(blue squares), and microlensing (dark red error-bar crosses) are shown.
The yellow shaded region indicates the
expected sensitivity of the Doppler method assuming a $15\,$year
survey with the present-day sensitivity. The cyan region indicates the expected sensitivity of the Kepler
mission. Black letters show the
locations of our Solar System's planets. The snow line is taken to
be at  $a_{\rm snow} = 2.7\,{\rm AU}\,M/M_\odot$.
\label{fig-m_vs_sep}}
\end{figure}

{\bf 3. Present-Day Microlensing Searches}

Microlensing searches today still basically carry out the approach advocated
by (Gould \& Loeb 1992): International networks of astronomers intensively
`follow-up' ongoing microlensing events that are discovered and `alerted' by two 
groups that search for events. 
Monitoring is done with 1m (and smaller) class telescopes.  Indeed,
because the most sensitive events are highly magnified, amateurs, with
telescopes as small as 0.25m, play a major role.

To date, 8 planets have been published, and one will be submitted soon
(See Fig 1).  All are beyond the snow line with equilibrium
temperatures $40\,{\rm K}<T<70\,{\rm K}$.  
Two are members of a
multi-planet system that resembles a solar system analog,
suggesting that such systems are probably not rare (Gaudi et al.\ 2008).
One is a low-mass planetary companion to a brown-dwarf star, which
suggests that such objects can form planetary systems similar to those
around solar-type main-sequence stars (Bennett et al. 2008).
Three are Jupiter class planets and so are similar to the planets found by RV at these
temperatures (Bond et al.\ 2004; Udalski et al. 2005; Dong et al. 2008b).  One of these
is the most massive planet ($\sim 3~M_{\rm Jup}$) yet found around
an M-dwarf (Dong et al. 2008a), whose existence may pose a challenge for the core-accretion theory of planet formation. 
Four are `super-Earth' planets,
with $3M_\oplus \la M \la 20 M_\oplus$, which are
substantially lighter than planets detected by RV at these
temperatures (Beaulieu et al. 2006; Gould et al.\ 2006; Bennett et al.\ 2008;
Sumi et al.\ 2009). These low-mass planet detections
imply that $\sim 20\%$ of stars have $\sim 10~M_\oplus$ planets with 
separations of $\sim 1.5-4~{\rm AU}$. 

\begin{figure}[ht*]
\epsscale{1.0}
\plottwo{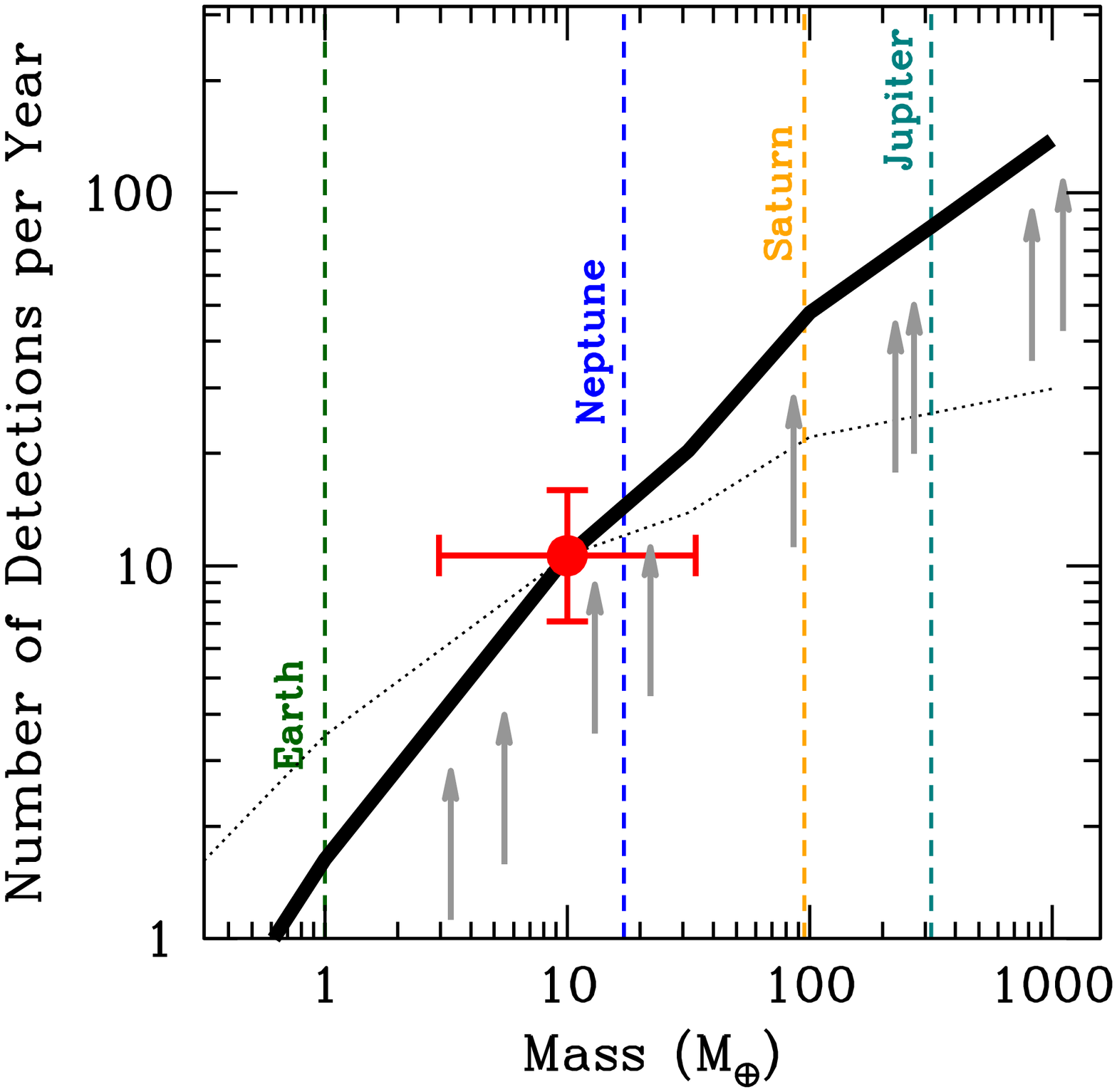}{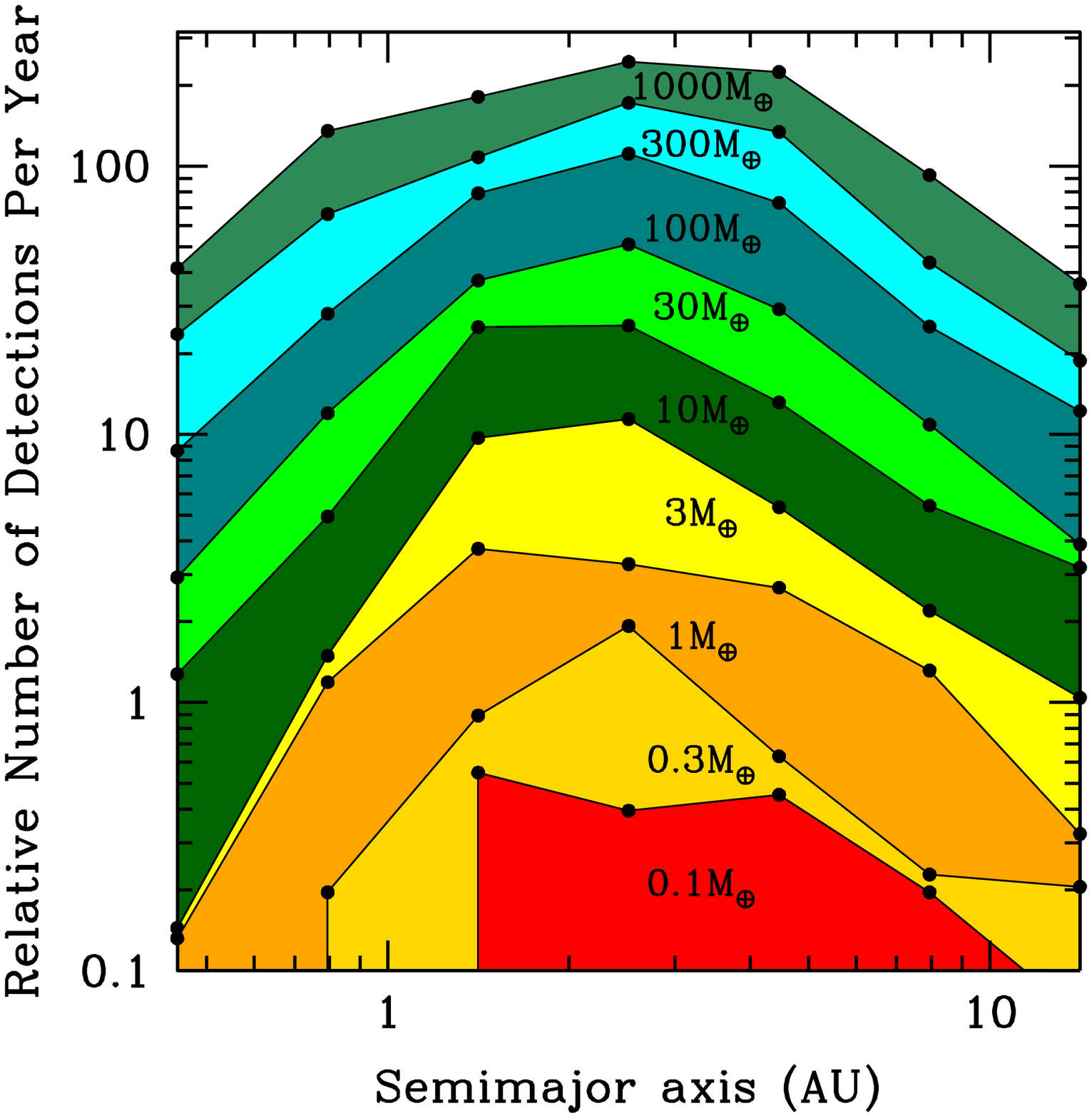}
\vskip-0.2in
\caption{
Expectations from the next generation microlensing survey,
including MOA-II, OGLE-IV, and a Korean telescope in South Africa.
(Left) Number of planets detected per year, normalized to 
number of $\sim 10M_\oplus$ found to date by microlensing (indicated by
the red dot). The solid line is the prediction assuming an equal number
of planets per logarithmic mass interval, and the dotted grey curve assumes
that the number of planets per log mass scales as $m^{-1/3}$. Planets are
assumed to be distributed
uniformly in $\log(a)$ between 0.4-20~AU. Arrows indicate the locations of
8 published exoplanets plus one unpublished planet (Sumi et al.\ 2009).
(Right) The semi-major axis distribution of the detected planets.
}
\end{figure}

{\bf 4. Next Generation Microlensing Searches}

While the early successes of the microlensing method are encouraging, it is
widely recognized that the rate
of microlensing detections cannot easily be increased beyond a few per year
without a change in strategy. 
Next-generation microlensing experiments will operate on completely
different principles from the current two-tier, `alert/follow-up' model.  
Instead, wide-field ($\sim 4\,\rm deg^2$) cameras on 1-2m telescopes on 3--4 continents will 
monitor large ($\sim 10\,\rm deg^2$) areas of the bulge once every 10 minutes
around-the-clock.  The higher cadence will more than double the number of events per
year, from $\sim 900$ to several thousand.  More important: {\it all of these events will 
automatically be monitored for planetary perturbations by the
search survey itself,} as opposed to roughly 50 events monitored per year as
at present.   

Detailed simulations of such a wide-FOV telescope network have been
performed by two independent groups (Bennett 2004; Gaudi, Gould, \&
Han, unpublished), both represented on this white paper. These simulations
assume 3 wide FOV, 1-2m telescopes located in Chile, Australia, and South
Africa.  Normalizing by the number of super-Earth ($\sim 10M_\oplus$) planets
found to date, and assuming a frequency of planets that is constant per log mass interval,
these simulations predict that this network will detect $\sim 1.6$ Earths and 10 super-Earths per year (see Fig.\ 2).
In particular, the Earth-mass planet detection rate will be higher than the 
current $\sim 1$ super-Earth per year detection
rate.  Finally, a next generation survey would
detect hundreds of Jupiter-mass free-floating planets per year if
every star has ejected a Jupiter-mass planet.

In fact, the microlensing community is close to acquiring the hardware needed
for such a next-generation survey.
The current microlensing survey groups, MOA and OGLE, have already
made significant progress in this direction.  MOA is currently operating the 1.8m MOA-II
telescope with a $2.2\, \rm deg^2$ FOV, and OGLE is in the midst of
upgrading the camera on its 1.3m telescope to the $1.4\, \rm deg^2$
for the OGLE-IV survey, which should commence in 2009.  
MOA also plans to upgrade to
a $10\,\rm deg^2$ camera within a few years. 
Finally, the South Korean congress has recently approved
the Korean Microlensing Telescope Network (KMTNet) project, which
will consist of three 1.6m-class telescopes with $\sim 4\,\rm deg^2$ CCD
cameras, to be built in South Africa, Chile, and Australia over the
next 5 years.

{\bf 4.1 US Role in Next Generation Surveys}

While it appears that the major telescope hardware investments for
the next generation planetary microlensing survey will be made almost
exclusively by other countries, including Japan, New Zealand, Poland,
and South Korea,
there are a number of avenues by which US can still play a
significant role in the future of ground-based microlensing planet
searches. It is essential that we take advantage of these
opportunities, as it is clear that this field is increasingly being
dominated by other countries, and we are in danger of losing our
current competitive edge when the next generation of students is
trained on next-generation experiments in other countries.  

Consider, for example, the KMTNet Project: this is an enormous undertaking, which
will likely benefit greatly from international participation.
Sharing of technical
expertise and experience between Korean and US scientists and
engineers with instrumentation expertise will likely be beneficial for
the timely and efficient design and construction of the three
facilities.  Furthermore, given the rather uncertain economic climate,
completion of this ambitious project may ultimately prove too costly for any
single government.  If the US can share some of the financial burden
through monetary and/or in-kind contribution, this would ensure the
project's success.  

Second, it may be advisable to augment the already-planned facilities
with additional telescopes.  For example, simulations indicate that
the addition of a fourth facility in Hawaii would increase the
detection rate by $\sim 20\%$, and would improve the characterization
of some poorly-sampled planetary perturbations. It may be possible to
use already-existing hardware for this purpose; for example the
Pan-STARRS PS1 telescope (Kaiser 2004) perfectly fulfills the
requirements for such a facility.

Third, US scientists will likely have an important role to
play in the development of modeling software to analyze the events
found in these surveys. There are currently only a handful of people in the
world that have the expertise required to model planetary microlensing
events, and a significant fraction of them are in the US.

Finally, there will be a substantial role for follow-up observations
in next-gen microlensing surveys.  These follow-up observations will
come in many varieties.  First, high-precision photometry of planetary
deviations alerted real-time will improve coverage and significance of
the deviations.  Second, next-gen
surveys will also dramatically improve prediction of high-mag events
prior to peak, which would improve planet detection by triggering more
frequent observations during peak planet detection sensitivity.  Just
as with current microlensing planet searches, many of these
observations could be carried out by small telescopes, such as those
of the Las Cumbres Observatory Global Telescope (Brown et al.\ 2006), 
and those belonging to amateur astronomers around the globe.  
Third, it will be necessary to characterize the host stars
using high resolution photometry with space-based or ground-based
adaptive optics facilities, such as is currently available with
VLT and Keck, and will be available with future facilities such as
JWST or future large telescopes such as TMT and GMT.

{\bf 6. Conclusion and Outlook}

Although microlensing searches have so far detected only a handful of
planets, these have {\it already} changed our understanding of planet
formation in the critical region beyond the snow line.  Next
generation microlensing surveys, which would be sensitive to tens of
``cold Earths'' in this region, are well advanced in design conception
and are starting initial practical implementation.  These surveys play
an additional crucial role as proving grounds for a space-based
microlensing survey, the results of which are likely to
revolutionize our understanding of planets over a broad range of
masses, separations, and host star masses (Bennett et al.\ 2009b).

Although the major telescope hardware investments for the next
generation planetary microlensing survey will be made almost
exclusively by other countries, the US can still play a significant
role, through sharing of technical expertise, analysis of
events from these surveys, follow-up observations, and potentially
monetary and/or in kind contributions to supplement or augment the
already-planned hardware investments.

{\bf References}
\small

Beaulieu, J.-P., et al. 2006, Nature, 439, 437\\
Bennett, D. P.\ 2004, ASP Conference Proceedings, 321, 59 (astro-ph/0404075)\\
Bennett, D. P., \& Rhie, S. H. 1996, ApJ, 472, 660\\
Bennett, D. P., \& Rhie, S. H. 2002, ApJ, 574, 985\\
Bennett, D.~P., Anderson, J., Bond, I.~A., Udalski, A., \& Gould, A.\ 2006, ApJL, 647, L171\\
Bennett, D. P., Anderson, J., \& Gaudi, B. S. 2007, ApJ, 660, 781\\
Bennett, D.~P., et al.\ 2008, ApJ, 684, 663\\
Bennett, D.~P.\ 2009a, in Exoplanets: Detection, Formation, Properties, Habitability (arXiv:0902.1761)
Bennett, D.~P., et al.\ 2009b, White Paper for the Astro2010 PSF Science Frontier Panel\\
Bond, I.~A., et al. 2004, ApJ, 606, L155\\
Brown, T.~M., et al.\ 2006, BAAS, 208, 5605\\
Dong, S., et al. 2008a, ApJ, in press (arXiv:0804.1354)\\
Dong, S., et al. 2008b, ApJ, submitted (arXiv:0809.2997)\\
Gaudi, B.~S., \& Gould, A. 1997, ApJ, 486, 85\\
Gaudi, B.~S., Naber, R. M., \& Sackett, P. D. 1998, ApJ, 502, L33\\
Gaudi, B.~S., et al. 2008, Science, 319, 927\\
Gould, A., \& Loeb, A. 1992, ApJ, 396, 104\\
Gould, A., et al. 2006, ApJ, 644, L37\\
Griest, K., \& Safizadeh, N. 1998, ApJ, 500, 37\\
Ida, S., \& Lin, D. N. C. 2004, ApJ, 604, 388\\
Kaiser, N.\ 2004, SPIE, 5489, 11\\
Lunine, J.~I., et al.\ 2008, Final Exoplanet Task Force Report (arXiv:0808.2754)\\
Mao, S., \& Paczynski, B. 1991, ApJ, 374, L37\\
Paczynski, B. 1986, ApJ, 304, 1\\
Sumi, T., et al.\ 2009, in preparation\\
Udalski, A., et al. 2005, ApJ, 628, L109
\end{document}